\documentclass[journal]{IEEEtran}

\usepackage{cite}

\usepackage{graphicx}

\ifCLASSINFOpdf

\else

\fi

\usepackage[cmex10]{amsmath}

\usepackage[tight,footnotesize]{subfigure}

\usepackage{url}

\begin{document}

\title{Device modeling of superconductor transition edge sensors based on the two-fluid theory}

\author{
    Tian-Shun~Wang,
    Guang-Can~Guo,
    Qing-Feng~Zhu,
    Jun-Xian~Wang,
        Tie-Fu~Li,
    Jian-She~Liu,
    Wei~Chen,
        and~Xingxiang~Zhou
\thanks{Work supported in part by the China National Natural Science
Foundation (grant No. 60836001) and the Central Government University
Fundamental Research Fund.}
\thanks{T-S. Wang, G-C. Guo, and X. Zhou are with the Department
of Optics and Optical Engineering, as well as the Key Laboratory of
Quantum Information, University of Science and Technology of China,
Hefei, Anhui, 230026, China (e-mail of X. Zhou: xizhou@ustc.edu.cn).
}
\thanks{Q-F. Zhu and J-X. Wang are with the Department of Astronomy,
University of Science and Technology of China,
Hefei, Anhui, 230026, China.}
\thanks{T-F. Li, J-S. Liu and W. Chen are with the Institute of
Microelectronics, Tsinghua University, Beijing, 100084, China.}
}


\maketitle

\begin{abstract}
In order to support the design and study of sophisticated large
scale transition edge sensor (TES) circuits, we use basic SPICE
elements to develop device models for TESs based on the
superfluid-normal fluid theory. In contrast to previous studies, our
device model is not limited to small signal simulation, and it
relies only on device parameters that have clear physical meaning
and can be easily measured. We integrate the device models in design
kits based on powerful EDA tools such as CADENCE and OrCAD, and use
them for versatile simulations of TES
circuits. Comparing our simulation results with published
experimental data, we find good agreement which suggests that device
models based on the two-fluid theory can be used to predict the
behavior of TES circuits reliably and hence they are valuable for
assisting the design of sophisticated TES circuits.

\end{abstract}


\IEEEpeerreviewmaketitle

\section{Introduction}

\IEEEPARstart{T}{he} past two decades have witnessed the rapid
development of the superconductor transition edge sensor technology
\cite{ref:Walton04,ref:Irwin05} and its successful application in a
wide range of scientific and instrumental fields
\cite{ref:Irwin96,ref:Cabrera00,ref:Krauss85,ref:Wollman00,ref:Tanakaa09}.
Most impressively, mid scale TES detector arrays with tens to
hundreds of pixels have been fabricated and deployed in Astronomy
telescopes \cite{ref:Ellis05,ref:Shirokoff09}. In the near future,
it is expected that much larger scale TES detector arrays,
potentially with thousands to tens of thousands of pixels, will
become available \cite{ref:Holland06}.

A fully functional TES detector array is a complex superconductor
circuit system because all TES sensors at the pixel level need
auxiliary supporting circuits for device biasing and signal readout.
As the scale of the detector array grows, more system level circuits
such as multiplexers
\cite{ref:Benford03,ref:Irwin02,ref:Chervenak99} become
indispensable too. It quickly becomes overwhelming to design and
integrate all the necessary devices and circuits when the system
size becomes large, and this challenge can only be met by elaborate
electronic design automation (EDA) tools specifically developed for
TES circuits.

Unfortunately, sophisticated tools that can support the simulation
and design of large scale TES circuits are unavailable presently. An
important reason for this deficiency is the lack of reliable TES
device models that can be integrated in existing EDA tools to
predict the behavior of TES circuits accurately. Since TESs are
highly nonlinear electrothermal devices, their behavior is
complicated and their modeling is difficult. Most previous research
is limited to small signal models \cite{ref:Irwin05} which cannot be
used for important tasks such as determining the required dc biases
and deriving the temperature sensitivity from easily measurable
device parameters. Some studies try to model the temperature
dependence of TES resistance using fitting functions such as the
hyperbolic function \cite{ref:Burney07}, the error function
\cite{ref:Miller01}, the Fermi function \cite{ref:Ukibe99} and other
expressions \cite{ref:Taralli09,ref:Roth00}. Though convenient in
producing resistance-temperature (R-T) curves matching experimental
data measured under certain conditions that are often very different
than the actual working conditions for the TES devices (see Section
\ref{subsection:R-T}), these models are not based on sound physical
considerations and their applicability is difficult to justify. More
seriously, since TESs are highly nonlinear devices, their R-T
dependence and electrical and thermal behavior are very sensitive
to how the circuits are designed and biased, as well as how the
system is operated and how the R-T curves are measured (see Section
\ref{subsection:R-T}). The fitting function approach that models the
TES resistance as a sole function of the device temperature cannot
capture this critical dependence on TES device's working conditions
and hence it is fundamentally flawed.

With the long term goal of making highly capable and integrated EDA tools that
can support the design and simulation of large scale TES circuits, in this work
we develop device models for TESs based on the superfluid-normal fluid
theory. We choose SPICE as the modeling tool and use only the most basic
SPICE circuit elements in order to be able to integrate our device model in
the widest possible variety of circuit simulators. With the two-fluid theory as
the underlying physical
mechanism, the device model has the advantage that it only relies on device
parameters that have clear physical meaning and can be measured easily.
After integrating the device models in design kits based on powerful EDA tools
such as CADENCE \cite{ref:cadence} and OrCAD \cite{ref:orcad}, we then use them
to perform a variety of simulations
of TES circuits and compare the results to published experimental
data to test the validity and accuracy of the device models.



\section{Device physics}

In this section, we elaborate on the device physics that our TES
model is based on. Since the TES sensor is an electrothermal device,
we divide our discussion into the electric and thermal properties of
the TES device.

\subsection{Electric behavior}
The functioning of a TES sensor relies on the sharp transition
between the device's superconducting and normal states which is a
very complex process. There are two well known theories to describe
the transition physics, the Skocpol-Beasley-Tinkham (SBT) model
\cite{ref:Skocpol74} based on the phase-slip events in type I
superconductors and the Kosterlitz-Thouless-Berezinsky (KTB) model
\cite{ref:Berezinskii72,ref:Kosterlitz73} based on flux vortex
creation and interaction in type II superconductors. Though all
superconductors used to fabricate TES devices are of type I, some authors
argue that in two-dimensional thin films the vortex model is
applicable \cite{ref:Fraser04}. The question which theory should be
used to build electronic models that can describe TES device's
behavior accurately, or whether either model is suitable for this
purpose at all, can only be answered by comparing the predicted
behavior with experimentally measured data.

In our work, we are interested in building a simple model that
captures the most important elements of the device physics and thus
can be easily used to simulate the behavior of the TES device with
reasonably good accuracy. For this purpose, we consider a simplified
two-fluid model \cite{ref:Irwin98} which has its root in the SBT
theory. In this model, the sensor current is separated into a
supercurrent $I_s$ and a normal current $I_n$. The total current is
then
\begin{equation}
 I = I_s+I_n,
\end{equation}
and a voltage $V$ can appear across the TES device because of the normal current.
According to the SBT theory, the supercurrent $I_s = C_I I_c(T)$, where
$I_c(T)$ is the temperature-dependent critical current of the TES film, and $C_I$
is the ratio of the time averaged critical current in the phase slip lines to
$I_c$. The normal current $I_n$ can be associated with the voltage across the
device by $I_n = V/(C_nR_n)$, where $R_n$ is the normal state resistance of the
TES device, and $C_n$ (usually approximately equal to 1) is the ratio of
the total resistance of the phase slip lines in the TES film to $R_n$.

In the two-fluid theory, the temperature dependence of the device's
critical current $I_c$ plays an essential role. In our simplified
device model, it is the underlying mechanism for the temperature
dependence of the TES resistance. For simple BCS superconductors
that behave in accordance with Ginzburg-Landau theory, we have
\begin{equation}
 I_s = I_{s0}(1-t)^{3/2},
\end{equation}
where $I_{s0}$ is the supercurrent of the TES device at 0
temperature and $t=T/T_c$ is the temperature normalized to the
device's critical temperature. For a single layer uniform film, the
critical current can be expressed as a function of the sample's
other parameters such as the heat capacity and normal resistance
\cite{ref:Irwin98}. Since most TES devices consist of multi-layer
films made of different metals and rely on the proximity effect
arising from such a structure, we do not expect this relation to
hold and the supercurrent $I_{s0}$ is an independent parameter in
our device model. Nonetheless, the supercurrent and its temperature
dependence can be easily measured.

Summarizing the main elements in the simplified two-fluid model, we can express the TES
device's equivalent resistance as
\begin{equation}
 R=\frac{V}{I_{s0}(1-\frac{T}{T_c})^{3/2}+V/R_n},
\label{eq:equiv-R}
\end{equation}
where $V$ is the voltage across the device. The nonlinear resistance
described in Eq. (\ref{eq:equiv-R}) is implemented in our device
model with the critical temperature $T_c$, supercurrent $I_{s0}$ and
normal resistance $R_n$ being independent device parameters. Though
highly simplified, it focuses on the most important mechanism
underlying the TES device and simulation results based on it are
consistent with many conclusions derived from
experimental data. Notice that we have assumed 0 applied magnetic
field. The effect of magnetic fields, as well as other factors that
can affect TES device's behavior, will be considered in improved
versions of the device model.

\subsection{Thermodynamics}
\begin{figure}
\centering
\includegraphics[width=2.5in]{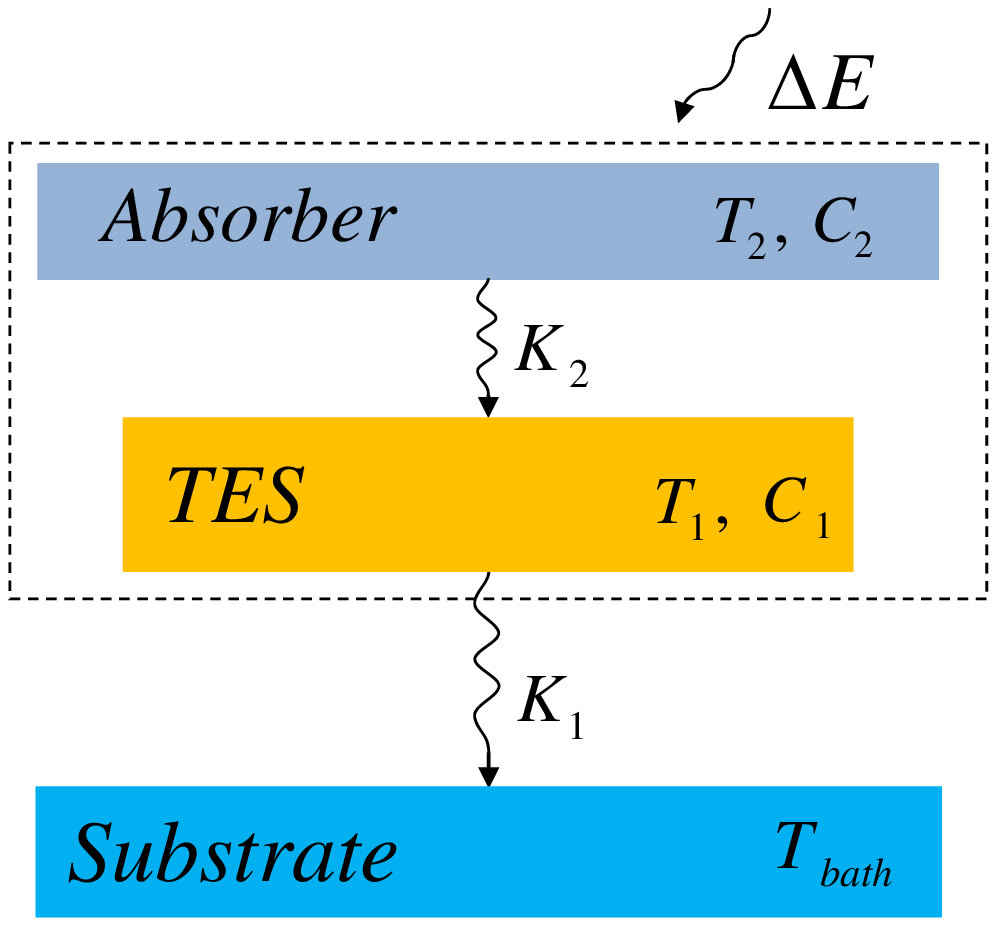}

\caption{Heat generation and conduction in the absorber-TES-substrate structure.}
\label{fig:3layer}
\end{figure}

The thermal behavior of the TES device is dictated by the interplay
of the Joule heating due to the device current and the heat
conduction to the substrate. To describe the involved physics, we
use a thermal model as shown in Fig. \ref{fig:3layer} which consists
of an absorber, the TES device, and the substrate. This model is
more comprehensive than most previous models which include only the
TES and substrate. Notice that, if we assign a very large value to
the absorber-TES heat conduction coefficient $K_2$, the heat
conduction between them is very efficient and they will remain at
the same temperature. Therefore, the thermal model in Fig.
\ref{fig:3layer} applies to devices without a dedicated absorber
too.

We assume that heat conduction between the absorber, TES and substrate are
governed by the power law
\begin{equation}
 P=K(T_a^n - T_b^n),
\end{equation}
where $P$ is the power flow between two elements $a$ and $b$, $T_a$ and $T_b$
are their temperatures, $K$ is the conduction coefficient, and $n$ is the
exponent. Assuming the substrate temperature is fixed at $T_{bath}$, we
can then write the thermal equation for the TES
\begin{equation}
 C_1\frac{dT_1}{dt} = IV-K_1(T_1^{n_1}-T_{bath}^{n_1})+K_2(T_2^{n_2}-T_1^{n_2}),
\label{eq:thermal-TES}
\end{equation}
where $T_1$ and $T_2$ are the temperatures of the TES and absorber, $C_1$ is the
heat capacity of the TES,
$I$ is the current through the TES, $V$ is the voltage across the TES,
and $K_1$, $K_2$, $n_1$ and $n_2$
characterize the TES-substrate and absorber-TES heat conduction. In Eq.
(\ref{eq:thermal-TES}), the terms on the right hand side correspond to
Joule heating and heat conduction to the substrate and from the absorber. Similar
consideration leads to the thermal equation for the absorber
\begin{equation}
 C_2\frac{dT_2}{dt} = P_s-K_2(T_2^{n_2}-T_1^{n_2}),
\label{eq:thermal-ABS}
\end{equation}
where $C_2$ is the heat capacity of the absorber and $P_s$ is the signal power.

Eqs. (\ref{eq:thermal-TES}) and (\ref{eq:thermal-ABS}) are the basis of the
thermal part of our device model which has $C_1$, $C_2$, $K_1$,
$K_2$, $n_1$ and $n_2$ as its independent parameters. These device parameters
have clear physical
meaning. Their values depend on the materials and geometries of the device and can
be measured by established experimental techniques. For simplicity, we have
neglected the temperature dependence of these device parameters which should
be weak in the temperature ranges that we are interested in.

\section{Modeling techniques}
The two options available for TES device modeling are SPICE and analog HDL (hardware
description language). In simulating and debugging TES circuits, we often need to
examine signals on the internal nodes of the TES device. Behavior models built with
analog HDL are less convenient for this purpose. Also, these models tend to be less
efficient in circuit simulation, and their integration in SPICE and SPICE-like
circuit simulators requires some effort. Because of these considerations, we choose
SPICE as our modeling tool.

Though we have simplified the device physics as much as possible in
section II, building SPICE models for the TES device is still quite
involved. The main challenge lies in constructing equivalent
electric circuit for the thermal part of the device model and
modeling the nonlinear elements and processes in the device. Many
latest circuit simulators have built-in nonlinear dependent source
support. However, the syntax is simulator specific and the
implementation details also vary. In order to be able to integrate
our device model in the widest possible variety of circuit
simulators, we choose to model the TES device using the polynomial
controlled source which is supported in almost all circuit
simulators.

The polynomial controlled source \cite{ref:url} is a circuit element
between two nodes whose voltage or current is dependent on one or
more controlling signals. In the element description, the number of
controlled signals, the nodes for the control signals, the
polynomial coefficients, and the initial conditions for the
controlling signals can be specified. For instance, a
voltage-controlled voltage source \texttt{Exx} between the positive
node \texttt{N+} and negative node \texttt{N-} can be described as
\texttt{Exx N+ N- POLY(ND) (NC1+ NC1-) ... P0 P1 ... IC=...}, where
\texttt{ND} is the number of dimensions (i.e. the number of
controlling signals), \texttt{NC1+, NC1- ...} are the positive and
negative nodes of the controlling signals, \texttt{P0, P1 ...} are
the polynomial coefficients, and the optional values following
\texttt{IC=} specify the initial conditions for the controlling
signals. Take as an example a two dimensional voltage source with
controlling signals $V_a$ and $V_b$, the controlled voltage $V_c$ is
\begin{equation}
 V_c = P_0 + P_1V_a + P_2V_b + P_3V_a^2 + P_4V_aV_b + P_5V_b^2 + ...
\end{equation}

\begin{figure}
  \centering
  \subfigure[Adder circuit between two voltages $V_{12}$ and
$V_{34}$. \texttt{E1} is the polynomial controlled source. The
output signal $V_{out}=V_{12}+V_{34}$. The resistances $R_{12}$,
$R_{34}$ and $R_{56}$ are chosen very large (e.g. 1G $\Omega$) to ensure
that the input and output resistances are large.]{
    \label{fig:macro-adder}
    \includegraphics[width=2.5in]{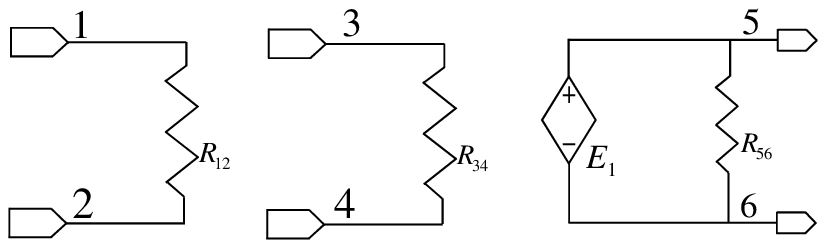}}
  \hspace{0.1in}
  \subfigure[Multiplier circuit between two voltages
$V_{12}$ and $V_{34}$. The output signal $V_{out}=V_{12}V_{34}$. $R_{12}$,
$R_{34}$ and $R_{56}$ are chosen to be very large.]{
    \label{fig:macro-multiplier}
    \includegraphics[width=2.5in]{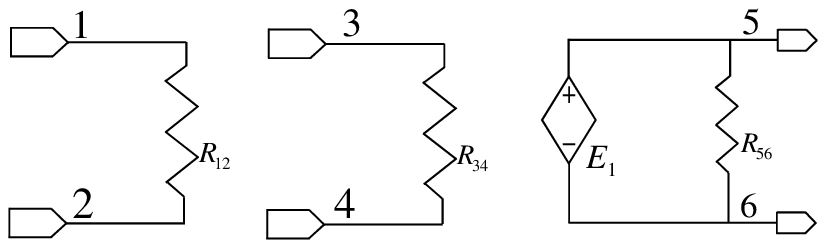}}
  \hspace{0.1in}
  \subfigure[Divider circuit for $V_{12}$ and
$V_{34}$. All resistances are large. The output signal
$V_{out}=V_{12}/V_{34}$.]{
    \label{fig:macro-divider}
    \includegraphics[width=2.5in]{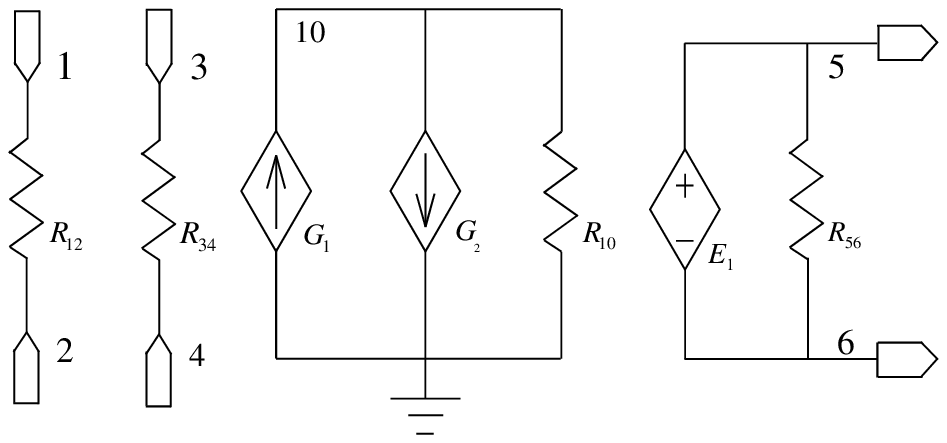}}
  \caption{Polynomial controlled source based circuits to add, multiply, and
divide two voltage signals. }
  \label{fig:macromodel}
\end{figure}

Seemingly simplistic, the polynomial controlled source is extremely
powerful and can be used to realize many operations on multiple
electric signals \cite{ref:Kielkowski95}. For example, the circuit
in Fig. \ref{fig:macro-adder} realizes the addition between two
voltages $V_{12}$ and $V_{34}$ with a polynomial controlled
source \\
\texttt{E1 5 6 POLY(2) (1 2) (3 4) 0 1 1} \\
To realize the multiplication between them,

use the polynomial controlled source
\\
\texttt{E1 5 6 POLY(2) (1 2) (3 4) 0 0 0 0 1} \\
instead, as shown in Fig. \ref{fig:macro-multiplier} .
For division
between two voltages $V_{12}$ and $V_{34}$, we use the circuit in
Fig. \ref{fig:macro-divider} where the two voltage controlled
current
sources \\
\texttt{G1 0 10 POLY(1) (1 2) 0 1} \\
and \\
\texttt{G2 10 0 POLY(2) (3 4) (10 0) 0 0 0 0 1} \\
play the central role. Since the currents in the two sources are
$I_{\texttt{G1}} = V_{12}$ and $I_{\texttt{G2}} = V_{34}V_{10}$ in value, we
have $V_{10} = (I_{\texttt{G1}}- I_{\texttt{G2}}) R_{10} = (V_{12}
-V_{34}V_{10}) R_{10}$. From this we can solve for the voltage across
nodes 10 and 0 which is
\begin{equation}
 V_{10} = \frac{V_{12}}{V_{34}+1/R_{10}}\approx \frac{V_{12}}{V_{34}}
\end{equation}
as long as the resistance $R_{10}$ is large. The voltage controlled voltage
source
\\
\texttt{E1 5 6 POLY(1) (10 0) 0 1} \\
in parallel with the large resistance $R_{56}$ simply mirrors the
voltage $V_{10}$ so that the output voltage $V_{56}$ is the division
between the input voltages $V_{12}$ and $V_{34}$.

In the following, we explain how the electric and thermal part of
the TES physics and the coupling and feedback between them are
modeled. We also describe relevant circuit diagrams.

\subsection{Electric behavior modeling}
\begin{figure}
  \centering
  \subfigure[Voltage controlled resistance between nodes 1 and 2. $R_x$ and
$R_1$ are 1$\Omega$. The input resistance $R_{in}$ is large.]{
    \label{fig:R_TES_circuit-vcres}
    \includegraphics[width=2.5in]{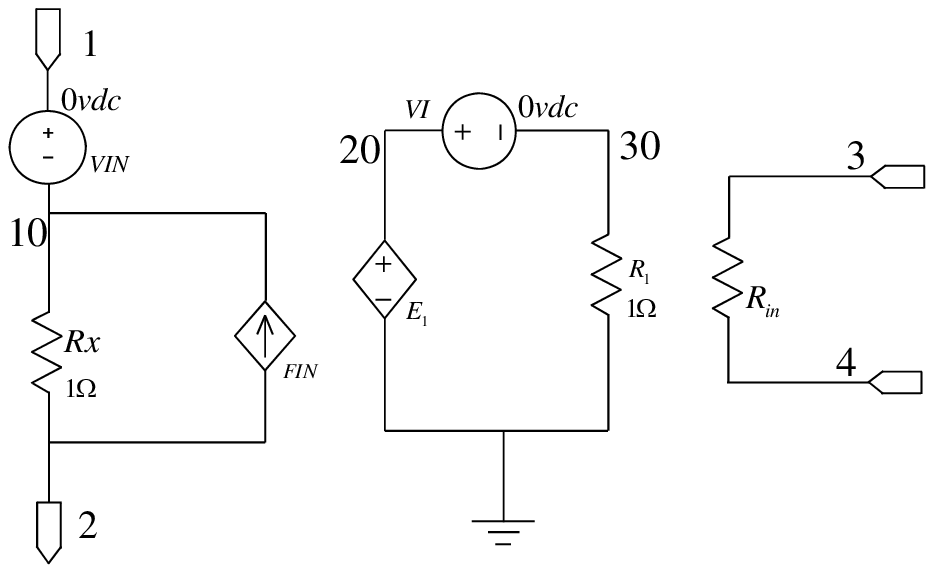}}
  \hspace{0.1in}
  \subfigure[Circuit to supply the input control voltage to the voltage
controlled resistance circuit in Fig. \ref{fig:R_TES_circuit-vcres}.
$R_{12}$ and $R_{10}$ are large. $V_{12}$ is the voltage across the
TES device. The adder circuit is that in Fig. \ref{fig:macro-adder}.
The divider circuit is shown in Fig. \ref{fig:macro-divider}. The
input signal $I_s$ is a voltage with a value equal to the
supercurrent of the TES. It is supplied by the circuit in Fig.
\ref{fig:SuperI_on_temp-supercurrent}. ]{
    \label{fig:R_TES_circuit-rtes}
    \includegraphics[width=2.5in]{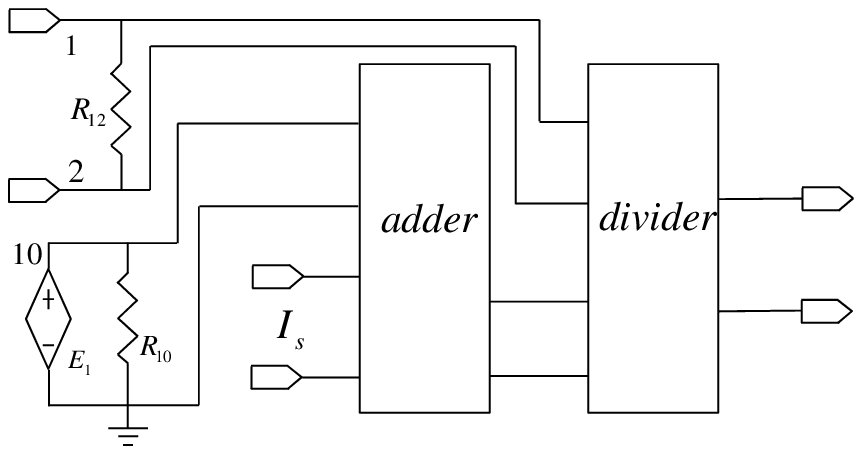}}
  \caption{Circuit model for the TES resistance.}
  \label{fig:R_TES_circuit}
\end{figure}

In order to model the voltage and temperature dependent TES resistance in Eq. (\ref{eq:equiv-R}),
we use the circuits shown in Fig. \ref{fig:R_TES_circuit}. At the heart of
the circuit is the effective voltage controlled resistance in Fig.
\ref{fig:R_TES_circuit-vcres} realized by the following polynomial
voltage controlled current source \\
\texttt{FIN 2 10 POLY(2) VIN VI 0 0 0 0 1} \\
The current in this controlled current source \texttt{FIN} is
\begin{equation}
I_{\texttt{FIN}} = I_{\texttt{VIN}}I_{\texttt{VI}},
\end{equation}
where $I_{\texttt{VIN}}$ and $I_{\texttt{VI}}$ are
currents in the auxiliary 0 voltage sources \texttt{VIN} and \texttt{VI}. $I_{\texttt{VI}}$ can
be calculated from the voltage of the voltage controlled voltage source \\
\texttt{E1 20 0 POLY(1) (3 4) -1 1} \\
and its value is ($\texttt{R1}=1\Omega$)
\begin{equation}
 I_{\texttt{VI}} = V_{\texttt{E1}}/\texttt{R1} = V_{34}-1
\end{equation}
where $V_{34}$ is the voltage across nodes 3 and 4. Since the total
current through the resistor \texttt{Rx} is
\begin{equation}
I_{\texttt{Rx}} = I_{\texttt{VIN}} + I_{\texttt{FIN}} =
I_{\texttt{VIN}}V_{34},
\end{equation}
the voltage across nodes 1 and 2 is $V_{12} = R_{x}I_{\texttt{Rx}} =
I_{\texttt{VIN}}R_{x}V_{34}$. The effective resistance between nodes
1 and 2 can then be calculated to be ($R_x = 1\Omega$)
\begin{equation}
 R_{12} = \frac{V_{12}}{I_{\texttt{VIN}}} = V_{34}.
\end{equation}
Notice that the effective resistance across terminals 1 and 2 is
controlled by the voltage $V_{34}$. If we design the circuit
appropriately so that $V_{34}$ is related to the voltage across
nodes 1 and 2 by the expression on the right hand side of Eq.
(\ref{eq:equiv-R}), we can then effectively realize a TES resistance
across these two nodes. This can be done by using the circuit in Fig.
\ref{fig:R_TES_circuit-rtes}. This circuit has two inputs, the TES voltage
$V_{12}$ and another voltage equal to the TES supercurrent $I_s$ in value. The
TES voltage is scaled by the voltage controlled voltage source \\
\texttt{E1 10 0 POLY(1) (1 2) 0 1/$R_{n}$} \\
and fed into the adder circuit which has $I_s$ as its other input.
The input signals to the divider circuit are the TES voltage and the
output from the adder circuit. The effective resistance across nodes
1 and 2 in Fig. \ref{fig:R_TES_circuit-vcres} is then
\begin{equation}
 R_{12} = \frac{V_{12}}{I_s + \frac{V_{12}}{R_n}},
\end{equation}
where $I_s$ is the supercurrent and $R_n$ is the normal resistance of the TES
device. The two-fluid theory based electric behavior of the TES device is then
successfully modeled by our circuit.

\subsection{Thermodynamics modeling}
SPICE is not designed to simulate thermodynamics. Though users can
specify a temperature in circuit simulation, it is a constant
ambient temperature used by device models to determine the electric
characteristics of circuit elements (e.g. the diode current depends
on not only its voltage bias but also the operation temperature). In
order to study how the temperature of the TES device depends on its
working condition, as well as how it changes in time, we must build
equivalent electric circuit to simulate its thermodynamics.

\begin{figure}
  \centering
  \subfigure[Equivalent electric circuit for the thermodynamics of the TES.
  $C_1$ and $C_2$ are the heat capacities of the TES and absorber. The voltage
$V_{Tbath}$ represents the environment temperature. Voltages on $C_1$ and $C_2$
represent the TES and absorber temperatures. $I_{P_J}$ and $I_{P_S}$ are the
Joule heat
  of the TES and the signal power. $I_{bath}$ and $I_{21}$ are the heat
conduction
  to the substrate and from the absorber.]{
    \label{fig:TES_thermal_circuit-thermal}
    \includegraphics[width=2.5in]{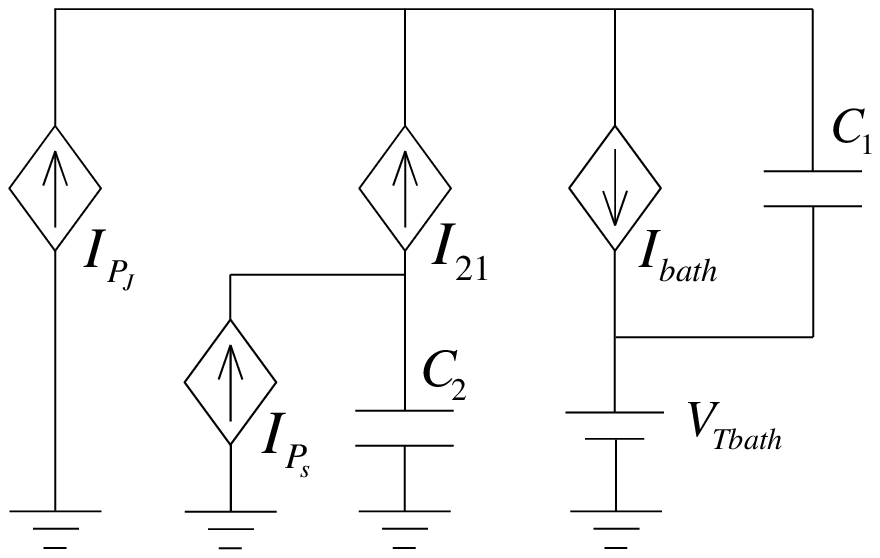}}
  \hspace{0.1in}
  \subfigure[Circuit to model the Joule heat in resistor $R_x$. The resistance
$R_1=1\Omega$.]{
    \label{fig:TES_thermal_circuit-joule}
    \includegraphics[width=2.5in]{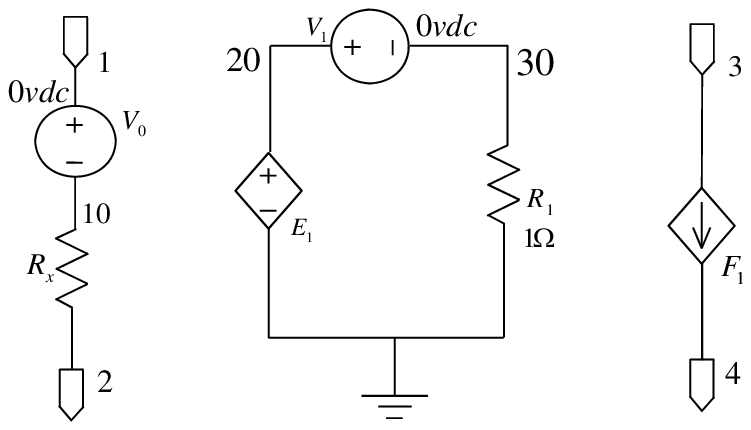}}
 \hspace{0.1in}
  \subfigure[Circuit to model the heat flow in the TES that obeys the power
law. $R_{12}$, $R_{34}$, $R_{10}$ and $R_{20}$ are large.]{
    \label{fig:TES_thermal_circuit-powerlaw}
    \includegraphics[width=2.5in]{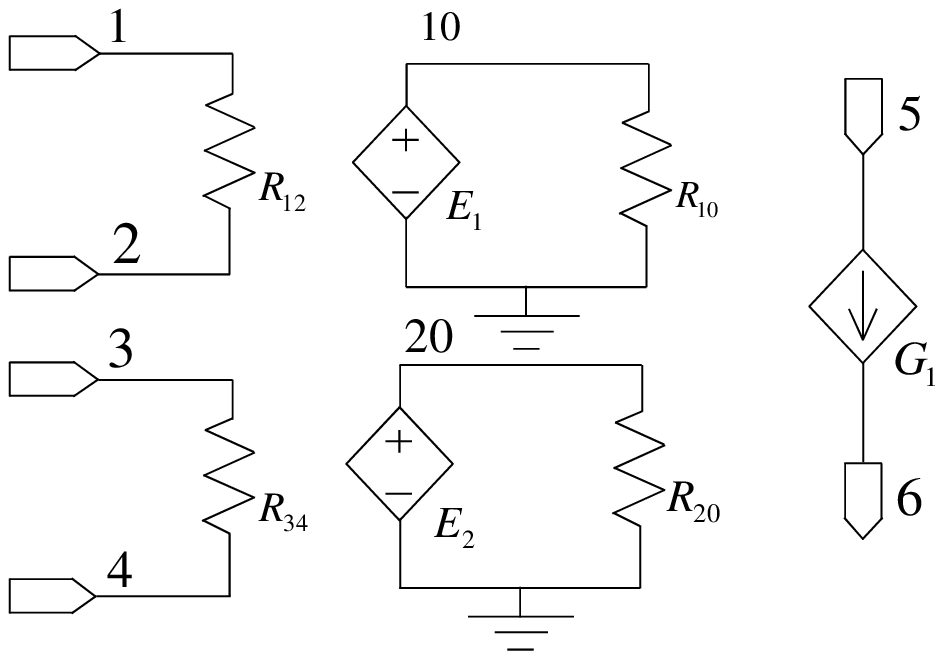}}
  \caption{Equivalent electric circuit for the thermal process in the
TES film.}
  \label{fig:TES_thermal_circuit}
\end{figure}

As shown in Fig. \ref{fig:TES_thermal_circuit-thermal} , the
thermal equation (\ref{eq:thermal-TES}) for the TES film can be
mapped to the electric equation of a capacitor being charged by
current sources whose values are given by the terms on the right
hand side of the equation. The voltage across the capacitor
corresponds to the temperature of the TES device, and the value of
the capacitance is the heat capacity of the device. The current
terms are dependent on the electric signals and temperature of the
TES, thus they can be modeled by controlled polynomial sources.

The circuit to model the Joule heating in equation
(\ref{eq:thermal-TES}) is shown in Fig.
\ref{fig:TES_thermal_circuit-joule} . In this circuit,
The voltage controlled voltage source \\
\texttt{E1 20 0 POLY(1) (10 2) 0 1} \\
simply duplicates the voltage $V_{\texttt{Rx}}$ across the resistor
\texttt{Rx} so that $V_{\texttt{E1}}=V_{\texttt{Rx}}$. The resistor
$\texttt{R1} = 1\Omega$ converts $V_{\texttt{E1}}$ to a current
$I_{\texttt{V1}}$ that is equal to $V_{\texttt{Rx}}$ in value:
\begin{equation}
 I_{\texttt{V1}} = V_{\texttt{E1}}/\texttt{R1} = V_{\texttt{Rx}}.
\end{equation}
Current in the current controlled polynomial current source \\
\texttt{F1 3 4 POLY(2) V0 V1 0 0 0 0 1} \\
is then
\begin{equation}
 I_{\texttt{F1}} = I_{\texttt{V0}}I_{\texttt{V1}} = I_{\texttt{Rx}}V_{\texttt{Rx}}
\end{equation}
which is equal to the Joule heat in the resistor \texttt{Rx}.
Using the TES device in place of \texttt{Rx}, we can then wire the
controlled current source \texttt{F1} in Fig.
\ref{fig:TES_thermal_circuit-thermal} to model the Joule
heat dissipated by the TES device.

The heat conduction terms on the right hand side of Eq.
(\ref{eq:thermal-TES}) can be directly modeled by controlled
polynomial sources. In Fig. \ref{fig:TES_thermal_circuit-powerlaw},
the voltage controlled polynomial voltage sources \\
\texttt{E1 10 0 POLY(1) (1 2) 0 0 0 0 0 1} \\
and \\
\texttt{E1 20 0 POLY(1) (3 4) 0 0 0 0 0 1} \\
produce two voltages $V_{10} = V_{12}^5$ and $V_{20} = V_{34}^5$, where
$V_{12}$ and $V_{34}$ represent temperatures of structures in the TES. The
polynomial controlled current source \\
\texttt{G1 5 6 POLY(2) (10 0) (20 0) 0 K -K} \\
then realizes a heat flow of $K(V_{12}^5 - V_{34}^5)$. The heat
conduction exponent $n_1$ and $n_2$ in Eq. (\ref{eq:thermal-TES})
are material dependent and a value of 4 or 5 are often used. If $n$
happens to be a non-integer (but rational) number, it can be written
as a fraction. From the numerator and denominator of the fraction,
we can construct appropriate power and root circuits using
polynomial controlled sources and realize the corresponding heat
flow.

Once we mapped the temperature of the TES to the voltage of a
capacitor, and modeled the Joule heat of the TES and its heat
flow to other parts of the system using polynomial controlled
sources, we can then use the circuit in Fig.
\ref{fig:TES_thermal_circuit-thermal} to describe the
thermal processes in the TES. The thermodynamics of the absorber in
Eq. (\ref{eq:thermal-ABS}) can be modeled using the same techniques.

\subsection{Electrothermal coupling and feedback}
The key to the operation of the TES device is the negative electrothermal feedback.
When the temperature of the TES rises due to the absorption of signal power, the
resistance of the device changes. This has the effect of changing the current and
Joule power of the device and its heat flow to the substrate which in turn regulates
the temperature of the device.

The effect of the TES resistance on the Joule power is already
modeled in the thermal circuit in Fig.
\ref{fig:TES_thermal_circuit-joule} where the equivalent
current source for the Joule heat is realized by a polynomial
current source controlled by the voltage and current of the TES.
When the nonlinear resistance of the TES device changes, so does its
Joule power.

\begin{figure}
  \centering
  \subfigure[Circuit to model the temperature dependent supercurrent of the
TES. $R_{12}$, $R_{10}$, $R_{20}$ and $R_{34}$ are large. The square
root circuit is shown in Fig. \ref{fig:SuperI_on_temp-sqrt}.]{
    \label{fig:SuperI_on_temp-supercurrent}
    \includegraphics[width=2.5in]{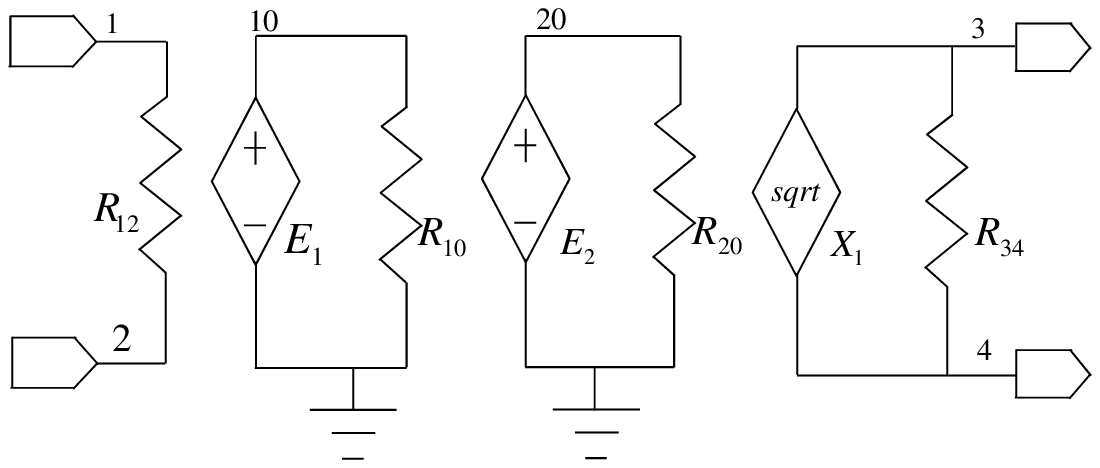}}
  \hspace{0.1in}
  \subfigure[The square root circuit. The divider circuit is shown in Fig.
\ref{fig:macro-divider}.]{
    \label{fig:SuperI_on_temp-sqrt}
    \includegraphics[width=2.5in]{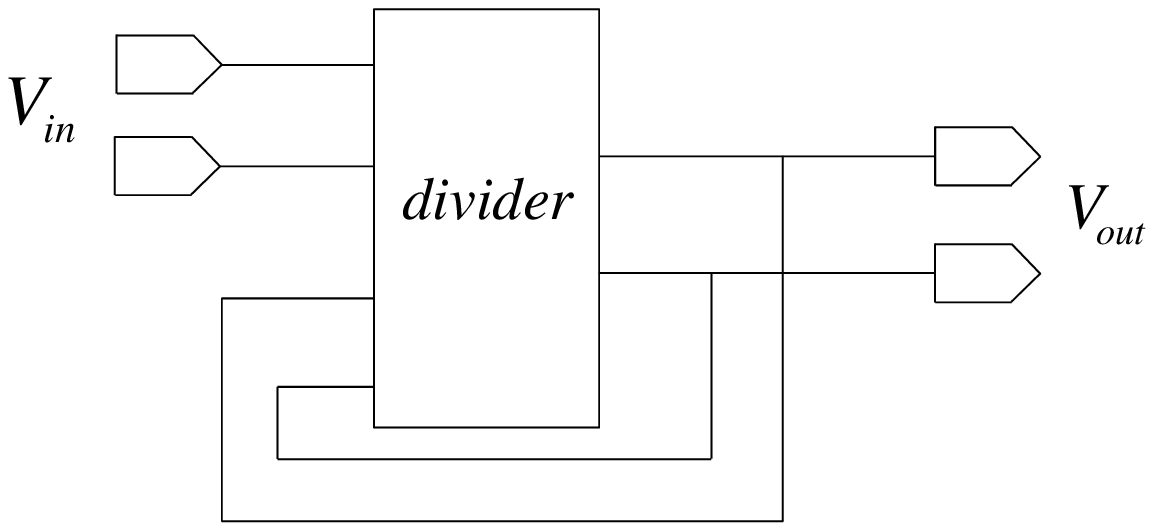}}
  \caption{Circuit to model the temperature dependence of the
supercurrent.}
  \label{fig:SuperI_on_temp}
\end{figure}

The effect of the TES temperature on the device's electric behavior
is manifested in the TES resistance in Eq. (\ref{eq:equiv-R}) where
the supercurrent changes with temperature. In order to model this
dependence, we use the circuit in Fig.
\ref{fig:SuperI_on_temp-supercurrent}. In this circuit, the input
voltage $V_{12}$ corresponds to the device
temperature $T$, and the voltage controlled voltage sources\\
\texttt{E1 10 0 POLY(1) (1 2) 1 -1/$T_c$} \\
and \\
\texttt{E2 20 0 POLY(1) (10 0) 0 0 0 $I_{s0}^2$} \\
in combination with the square root circuit \\
\texttt{X1 20 0 3 4  sqrt} \\
produce an output signal
\begin{equation}
 I_{s0}(1-\frac{T}{T_c})^{3/2}.
\end{equation}
This is the temperature dependent supercurrent $I_s$ of the TES device, and it
is fed into the circuit in Fig. \ref{fig:R_TES_circuit-rtes} to model the TES
resistance. The square root circuit is based on the divider circuit as shown in
Fig. \ref{fig:SuperI_on_temp-sqrt} where one of the input voltages
to the divider circuit is set to the output. Since $V_{out}=
V_{in}/V_{out}$, we have $V_{out} = \sqrt{V_{in}}$.

Once we have designed circuits to model the electric and thermal
behavior and the electrothermal feedback, we can construct a
complete device model for the TES based on them. Notice that our TES
device model is general purpose and can be used for important
studies not supported by the small signal models developed in
previous work.

\section{Simulation based on the device model}
Now that we have built the TES device model, we are interested in using it for
simulation of TES circuits to test its validity. Considering the simplicity of
the model and the large number of poorly understood and controlled factors in
TES device fabrication, it is unrealistic to expect that simulation results
based on our model will numerically agree with experimental data to exceedingly
high precision for every
fabricated TES device. However, a correct device model should give results that
are consistent with important qualitative conclusions drawn from
experimental data. By doing circuit simulation, we can also perform critical
research on TES circuit design and operation. This includes important studies
not possible before when only small signal models were available, such as
determining the optimal bath temperature and electrical bias points for TES
circuit operation and finding the allowed parameter margin space for TES device
fabrication.

For the purpose of circuit simulation, we integrate the device model in popular
EDA tools and leverage the power of these tools to carry out our studies. We use
CADENCE and OrCAD which are based on UNIX and WINDOWS platforms respectively.
The integration process mainly involves constructing subcircuits used in the
device model,
creating symbol views for them, and building a component library which contains the
necessary subcircuits and the TES device model circuit itself. Once all subcircuits
and components are created and tested, we can then use the graphic user interface
(GUI) provided by the EDA tools to draw TES circuits, specify device parameters and
run simulations. TES devices can be dragged into a circuit schematic and wired up to
the rest of the circuit just
like any other circuit elements such as resistors and inductors, and the EDA tools
will automatically generate the circuit netlists, add the stimulus and device models
and run the simulation using a simulator specified by the user. This
greatly improves the efficiency of our research and reduces human error.

In the following, we describe some interesting TES circuit simulations we
performed. We also analyze the results and check them against published
experimental observations when possible.

\subsection{Resistance-temperature (R-T) dependence}
\label{subsection:R-T}
The width of the superconducting to normal
transition is an important characteristic of the TES because the
sharpness of the transition determines its temperature sensitivity.
Some authors have tried to model the TES resistance $R$ using
fitting functions that give the measured transition width $\Delta T$
and normal state resistance $R_n$. Examples include the hyperbolic
function (with the empirical parameter $b$) \cite{ref:Burney07}
\begin{equation}
 R(T) = \frac{R_n}{2}\{tanh(\frac{T-T_c}{\Delta T/b})+1\},
\label{eq:Fitting_tan}
\end{equation}
the error function \cite{ref:Miller01}
\begin{equation}
 R(T) = \frac{R_n}{2}\{erf(\frac{T-T_c}{2\Delta T})+1\},
\label{eq:Fitting_err}
\end{equation}
as well as other mathematical expressions \cite{ref:Ukibe99,
ref:Taralli09,ref:Roth00}. One
notable problem with the fitting function approach is that the
temperature sensitivity calculated from the derivative of the
measured R-T curve is often much larger than that inferred from the
device's temporary response to a signal pulse. This is because the
fitting function approach completely ignores the dependence of the
R-T curve on the device's working conditions which is often
critical. We can study this issue using our device model.

\begin{figure}
\centering
\includegraphics[width=2.0in]{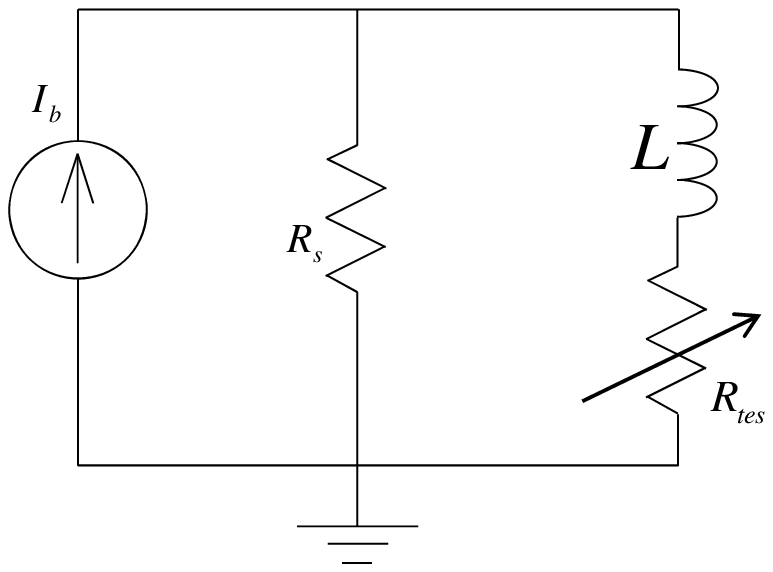}

\caption{Voltage biased TES device. $R_s$ is a small shunt
resistance and $I_b$ is the bias current.} \label{fig:TES_Vbiased}
\end{figure}

\begin{figure}
  \centering
  \subfigure[Simulated R-T curve for different bias
current of the TES device.]{
    \label{fig:RT_curve:rt}
    \includegraphics[width=2.5in]{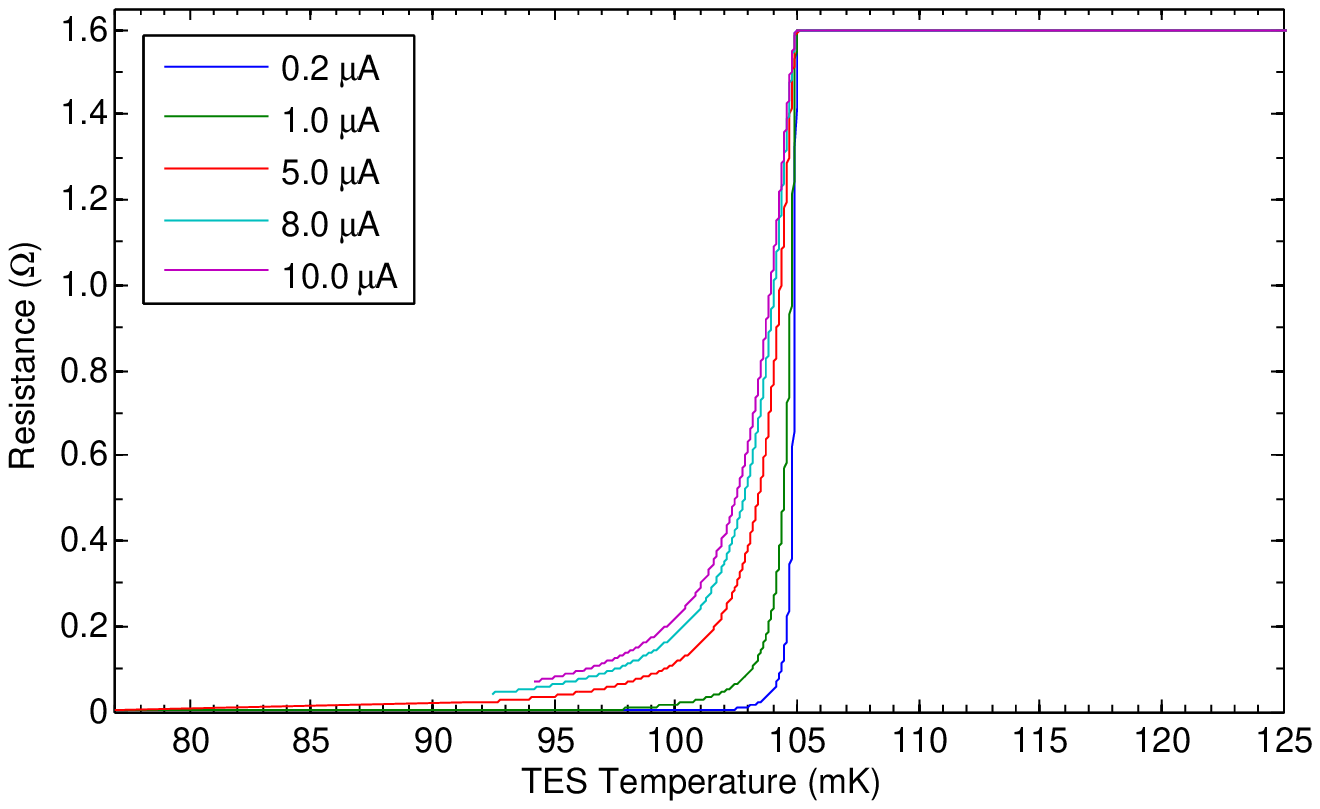}}
  \hspace{0.1in}
  \subfigure[The TES sample temperature curve.]{
    \label{fig:RT_curve:TBATH}
    \includegraphics[width=2.5in]{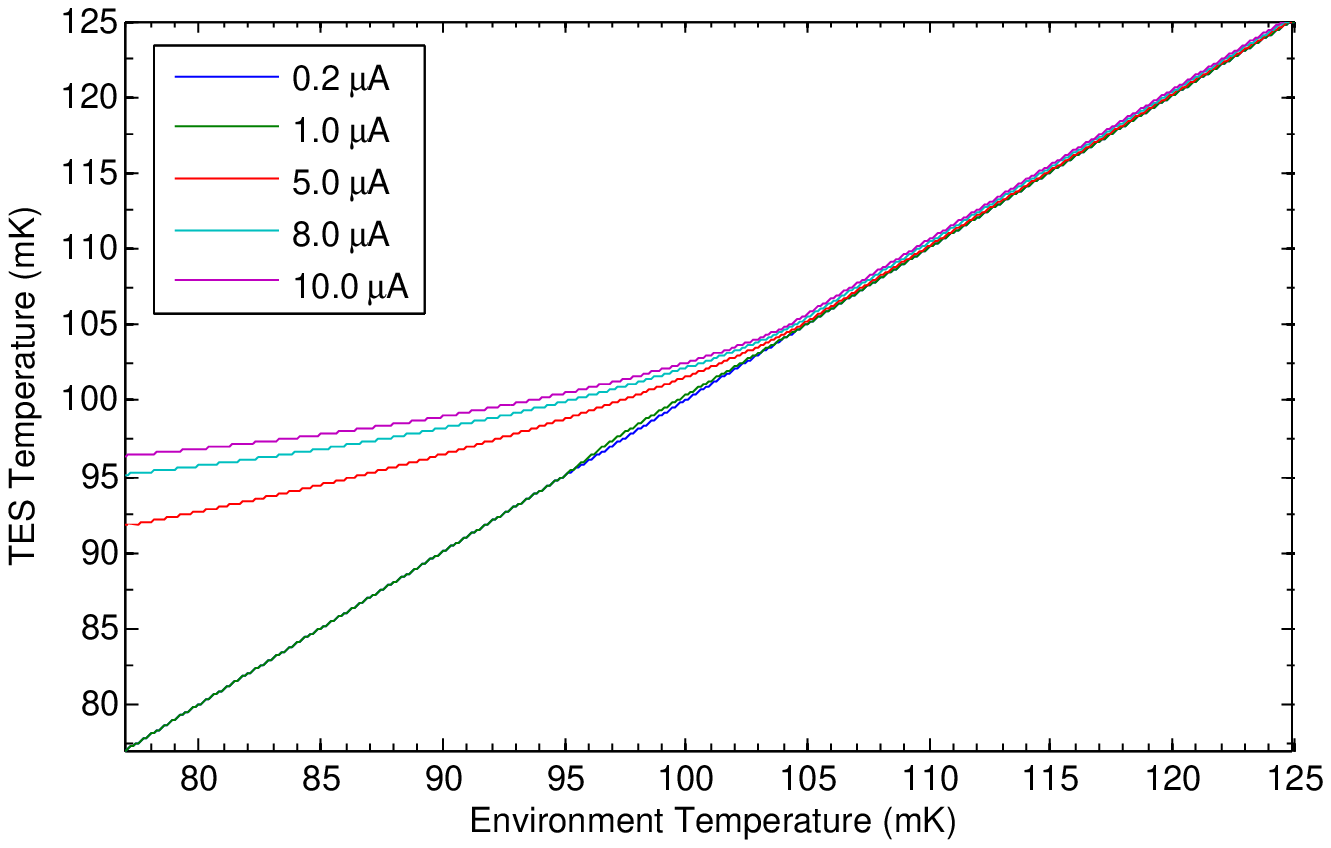}}
  \caption{Simulation of the R-T curve for TES under different bias currents.
The device parameters are taken from reference \cite{ref:Taralli09}.
The critical temperature of the TES is $T_c = 105$mK. The TES heat
capacity $C=3.3f$J/K. The shunt resistance $R_s=9.5m\Omega$. The
normal state resistance $R_n=1.6\Omega$. The heat conduction
coefficient $K=13.54$$W/K^5$. The 0 temperature supercurrent
$I_{s0}$ is estimated to be $35\mu$A.}
  \label{fig:RT_curve}
\end{figure}

The circuit for a voltage biased TES device is shown in Fig.
\ref{fig:TES_Vbiased}. The R-T curves are usually measured by
biasing the TES sample with a constant near 0 current and sweeping
the sample temperature. The reason to use a very small bias current
is to minimize the Joule heat so that the TES sample remains at the
same temperature with the substrate and environment. This
temperature can be set and changed by the temperature controller of
the refrigerator system, thus making it possible to measure the
sample resistance at different temperatures. We can simulate this
measurement process by a DC analysis based on our device model in
which the environment temperature $T_{bath}$ is the sweeping
parameter. For this simulation, we need an exhaustive set of TES device
parameters which is unfortunately not given in most published works.
We use the data from reference \cite{ref:Taralli09} which is
relatively complete. The result of the simulation is shown in Fig.
\ref{fig:RT_curve:rt}. In order to check that the TES sample remains
at the same temperature with the environment, the TES temperature is
plotted against the environment temperature in Fig.
\ref{fig:RT_curve:TBATH}. As can be seen from the figures, even
though in the transition region the current biases are small enough
to produce negligible Joule heat so that the TES sample remains at
the same temperature with the environment, the transition width
under each bias current can be quite different. Generally speaking,
the smaller the bias current, the sharper the transition. This
result clearly indicates that it is fundamentally flawed to model
the TES resistance using fitting functions like those in Eqs.
(\ref{eq:Fitting_tan}) and (\ref{eq:Fitting_err}) without specifying
the bias current under which the R-T curve is measured.

\begin{figure}
\centering
\includegraphics[width=2.5in]{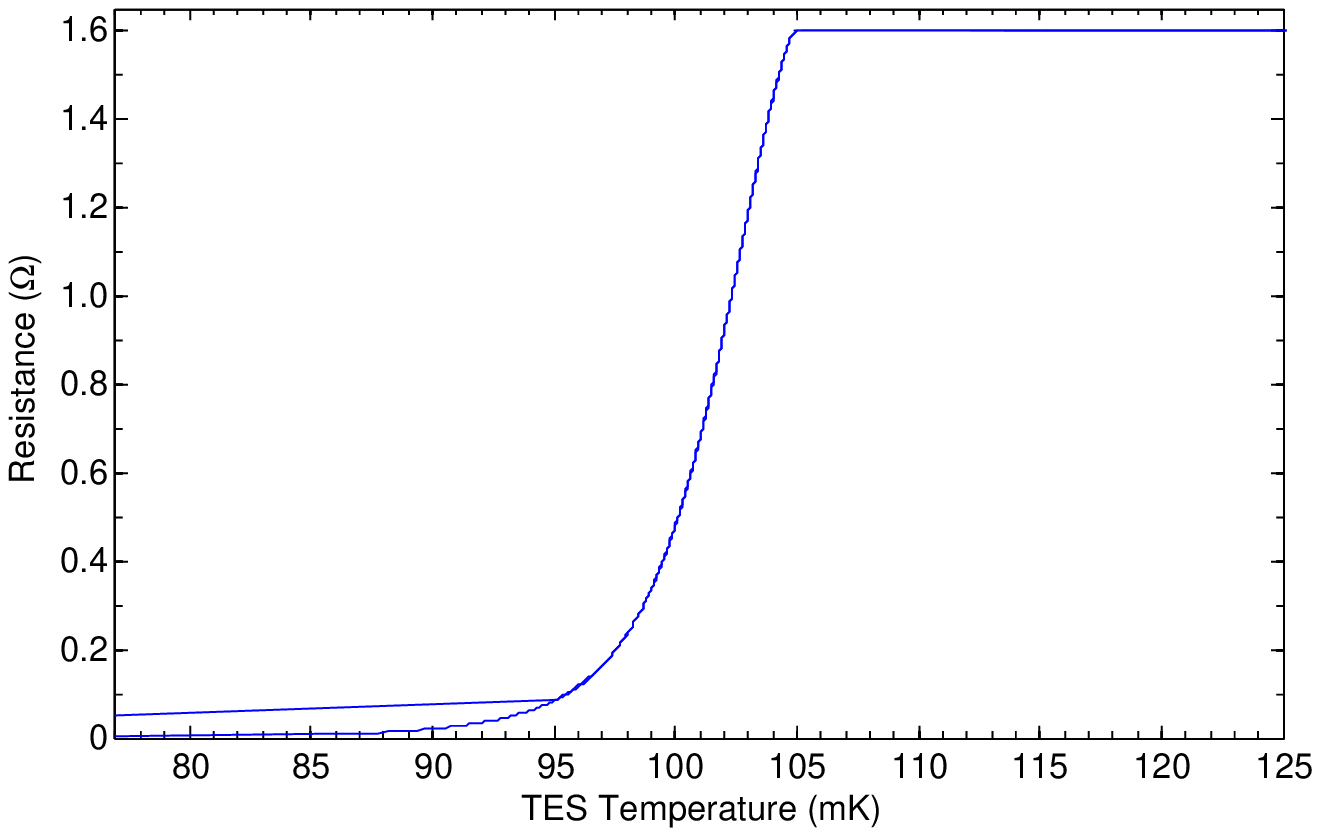}

\caption{The resistance-temperature curve of the TES in Fig.
\ref{fig:TES_Vbiased} produced by sweeping the bias current $I_b$.
The environment temperature is fixed at $T_{bath}=55m$K. The device
parameters are the same with those in Fig. \ref{fig:RT_curve}.}
\label{fig:RT_Vbiased}
\end{figure}

The working condition of the TES device is different than that for R-T curve
measurement. The environment temperature is set below the device's critical
temperature, and a nonzero bias current is applied to bring the device's
temperature to within the transition region. To determine the device's R-T
dependence under this working condition, we perform a dc analysis in which
$T_{bath}$ is fixed and the circuit's bias current $I_b$ in Fig.
\ref{fig:TES_Vbiased} is swept. The TES resistance is plotted against the device
temperature in Fig. \ref{fig:RT_Vbiased}.

Comparing the results in Fig. \ref{fig:RT_curve} and Fig. \ref{fig:RT_Vbiased},
we notice that the transition width of a working TES device is much
wider than that measured with near 0 bias current. While the transition width
measured with near 0 current can be as low as sub mili Kelvin, the value for a
working TES device is a few mili Kelvins. This explains why the TES device's
temperature sensitivity
\begin{equation}
 \alpha = \frac{T}{R}\frac{\partial\log{R}}{\partial\log{T}}
\end{equation}
($T$ is the operation temperature of the TES and $R$ the resistance
at $T$) calculated from the derivative of the measured R-T curve is
usually much greater than that inferred from the device's transient
response to an input signal. The R-T dependence in Fig. \ref{fig:RT_Vbiased}
cannot be easily verified
experimentally since the temperature of a working TES device cannot
be measured directly, and our simulation allows to study it in
detail. In setting the working condition for the TES device, it is
nontrivial to determine values for the substrate temperature and
bias current to optimize the device's temperature sensitivity and
other critical characteristics. Circuit simulations can help greatly
in finding appropriate bias points and working conditions for the
TES device. Otherwise, large number of measurements of the circuit's
IV characteristics must be performed.

\subsection{Hysteresis in the temperature-bias current curve}

Another interesting phenomenon of the TES device is that it can display
hysteresis when its temperature is adjusted with a bias current. In Fig.
\ref{fig:TES_Vbiased}, when the bias current is increased, the temperature of
the TES devices rises above the substrate temperature (i.e. the environment
temperature) because of the Joule heating of the TES. Though the temperature -
bias current curve cannot be directly measured (due to the difficulty in
measuring the temperature of the TES), the characteristics of this curve have
profound impact on the operation of the TES device and are therefore worth
careful investigation. A DC analysis based on our device model can be used for
this study.

\begin{figure}
  \centering
  \subfigure[Hysteretic temperature - bias current curve for the
TES device. The 0 temperature supercurrent $I_{s0} = 35\mu A$. Other device
parameters are the same with those in Fig. \ref{fig:RT_curve}.
Environment temperature is $55m$K.]{
    \label{fig:T_Ib-1} 
    \includegraphics[width=2.5in]{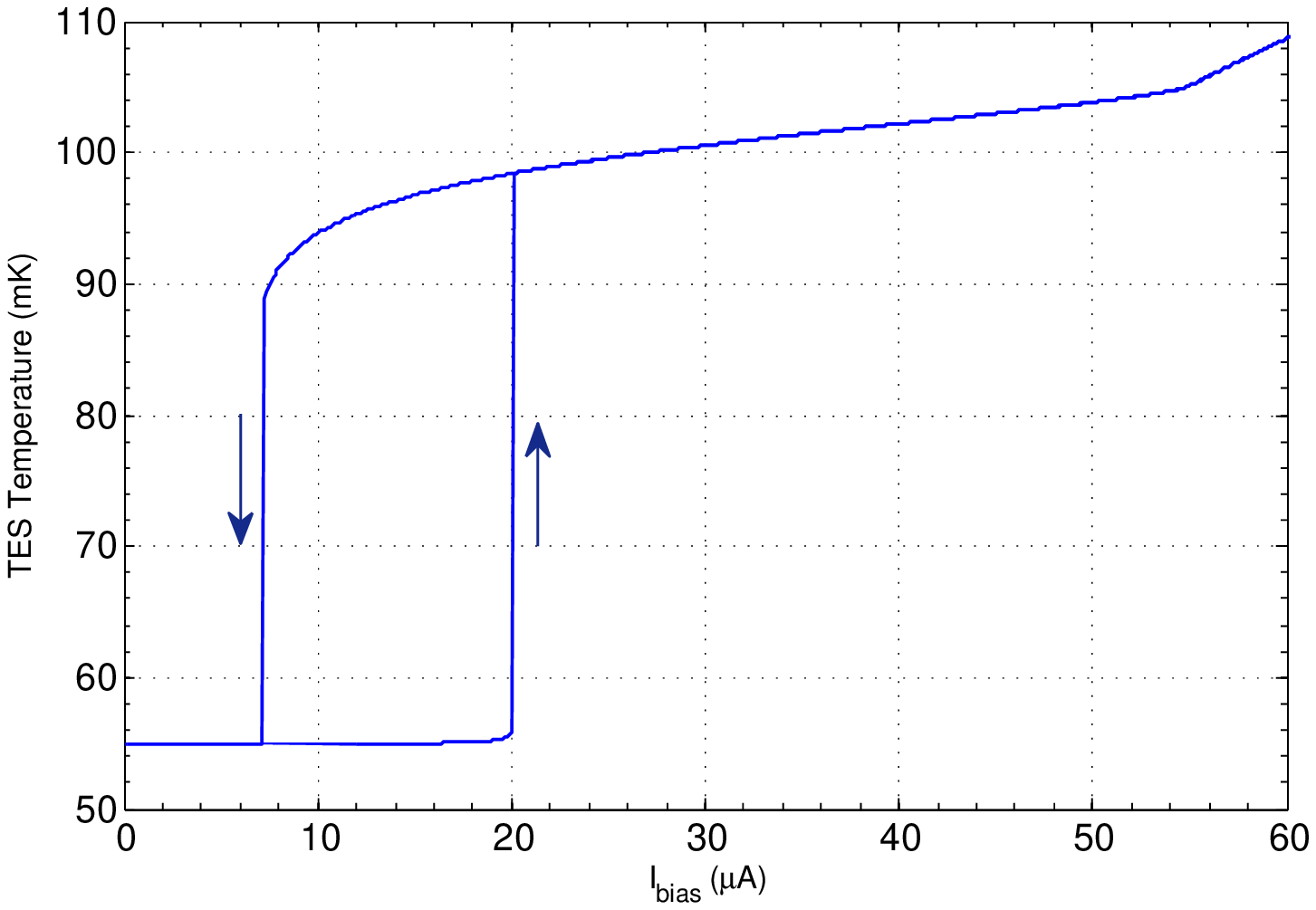}}
  \hspace{0.1in}
  \subfigure[Non-hysteretic temperature - bias current curve. $I_{s0}
  =3.9\mu A$, other parameters are the same as in Fig. \ref{fig:T_Ib-1}.]{
    \label{fig:T_Ib-2} 
    \includegraphics[width=2.5in]{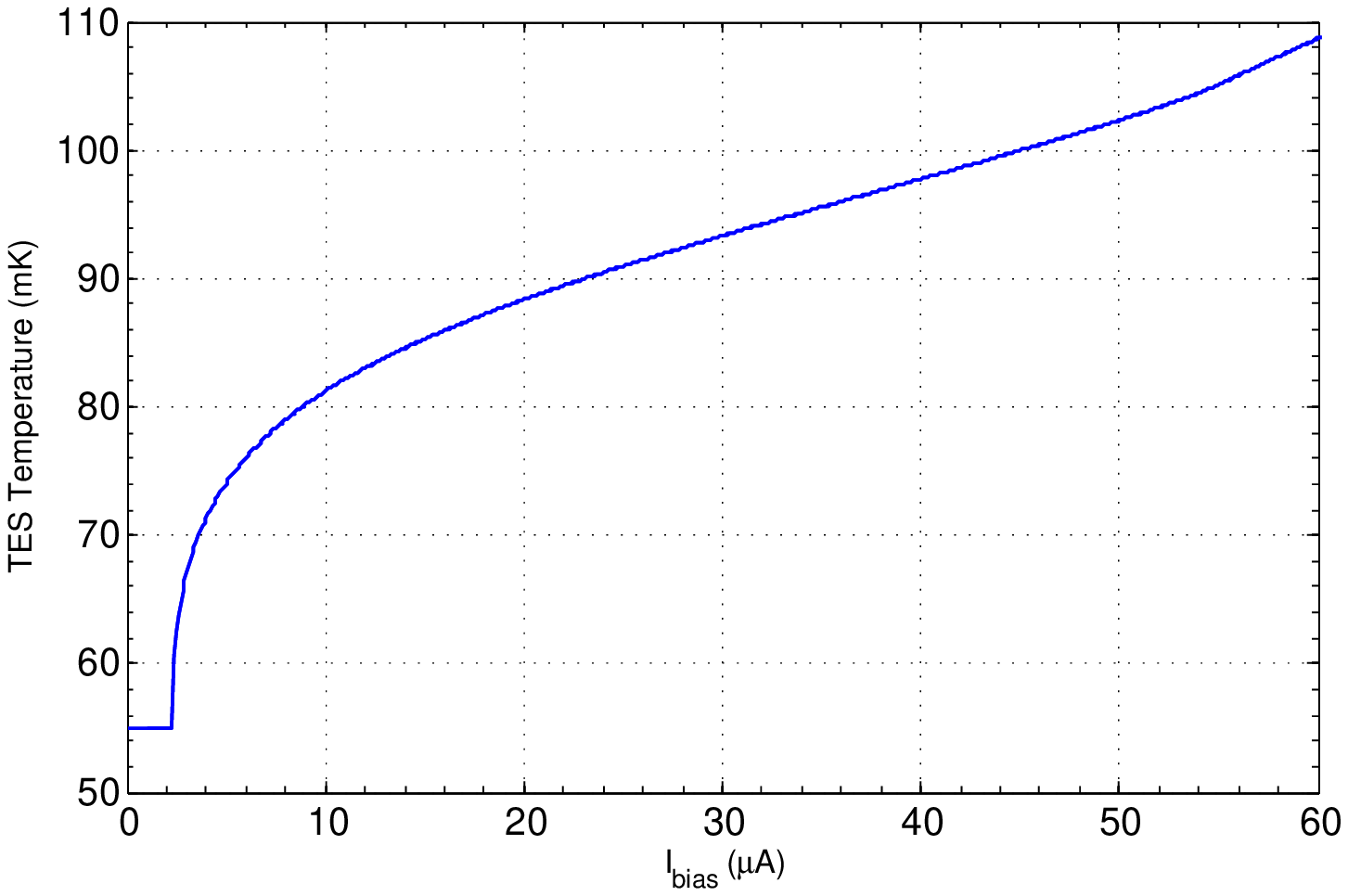}}
  \caption{Simulated temperature - bias current curve for the TES for
different device parameters.}
  \label{fig:T_Ib} 
\end{figure}

In Fig. \ref{fig:T_Ib}, the simulated temperature - bias current curve of
voltage biased TES device with different parameters are plotted. It can be seen
that when the bias current is increased, the temperature of the TES does not
increase with the bias current linearly. Instead, at some bias point it makes a
sharp transition from a value close to the substrate temperature to a value
close to the critical temperature of the device. More interestingly, for many
device parameters, this sudden transition between near substrate temperature
and near critical temperature can exhibit a hysteresis.
After the TES temperature has made a sudden transition to close to
the critical temperature at some bias current $I_{b1}$, if we subsequently
decrease the bias current we can bring the device temperature back to close to
the substrate temperature. This later temperature transition occurs suddenly
too at some bias current $I_{b2}$, and $I_{b2}$ can be different than $I_{b1}$
giving rise to the hysteresis shown in Fig. \ref{fig:T_Ib-1}.

The hysteresis in Fig. \ref{fig:T_Ib-1} is a consequence of the nonlinear nature
of the TES device. Simulations show that for certain device parameter ranges
the sudden temperature transition points can be very close to the critical
temperature of the device and this can disrupt the normal operation of the TES
and reduce its saturation input energy.
In order to avoid such a scenario, care must be taken in the design phase to
choose the device parameters correctly before it is fabricated. Such design
work relies on large number of simulations of the circuit behavior and
appropriate TES device models are indispensable.

\begin{figure}
  \centering
  \subfigure[Experimentally measured current in the TES against the total bias
current in Fig. \ref{fig:TES_Vbiased}. The curve is taken from reference
\cite{ref:Taralli09}.]{
    \label{fig:Ttes_Ib-experiment}
    \includegraphics[width=2.5in]{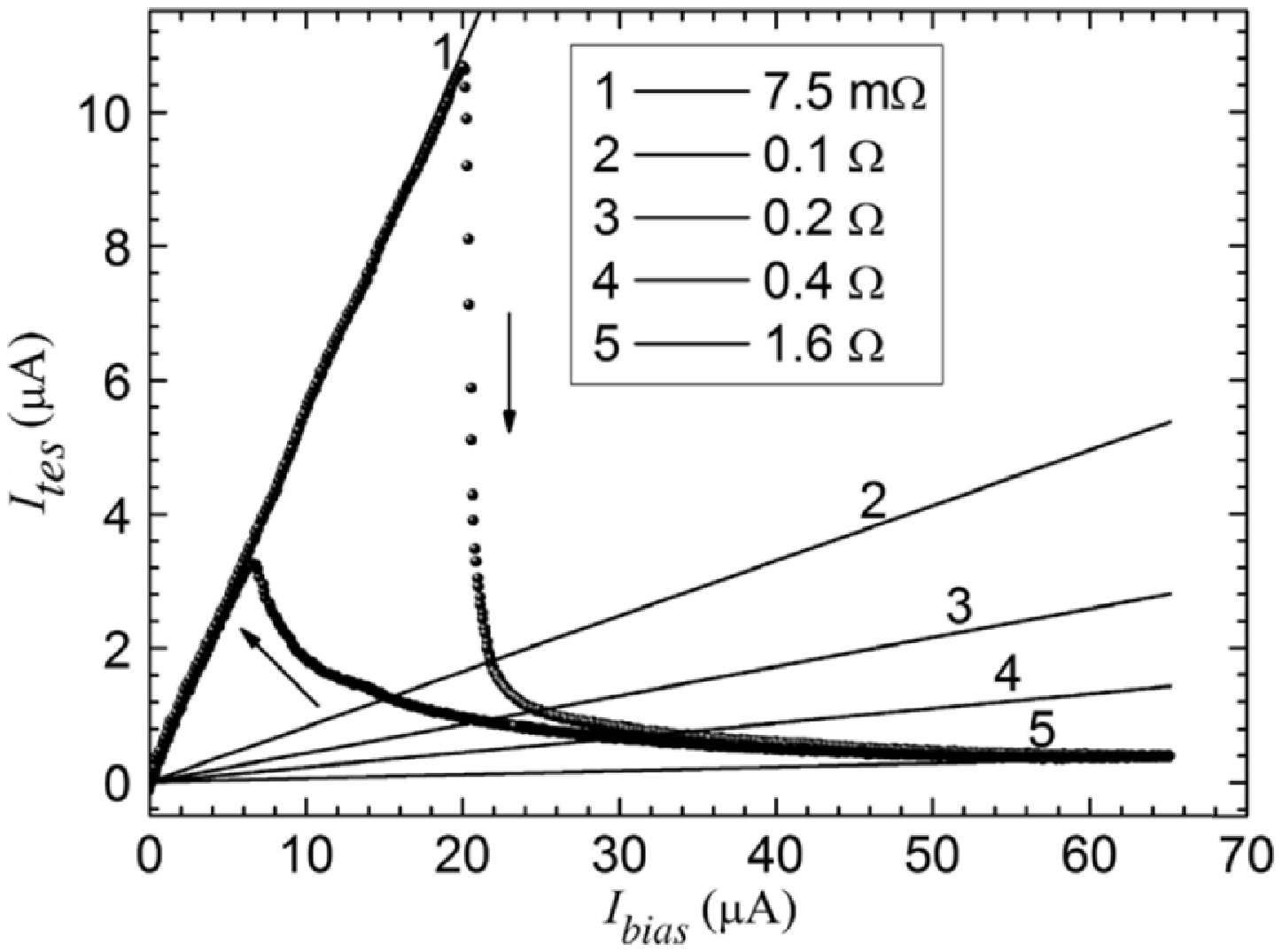}}
  \hspace{0.1in}
  \subfigure[Simulated $I_{tes}-I_{bias}$ curve. Device parameters given in
reference \cite{ref:Taralli09} are used. (See Fig. \ref{fig:RT_curve}.) The
parasitic resistance for the TES is estimated to be $8m\Omega$ according to the
superconducting branch in Fig. \ref{fig:Ttes_Ib-experiment}. $I_{s0}$ is
estimated to be $35\mu$A.]{
    \label{fig:Ttes_Ib-simulation}
    \includegraphics[width=2.5in]{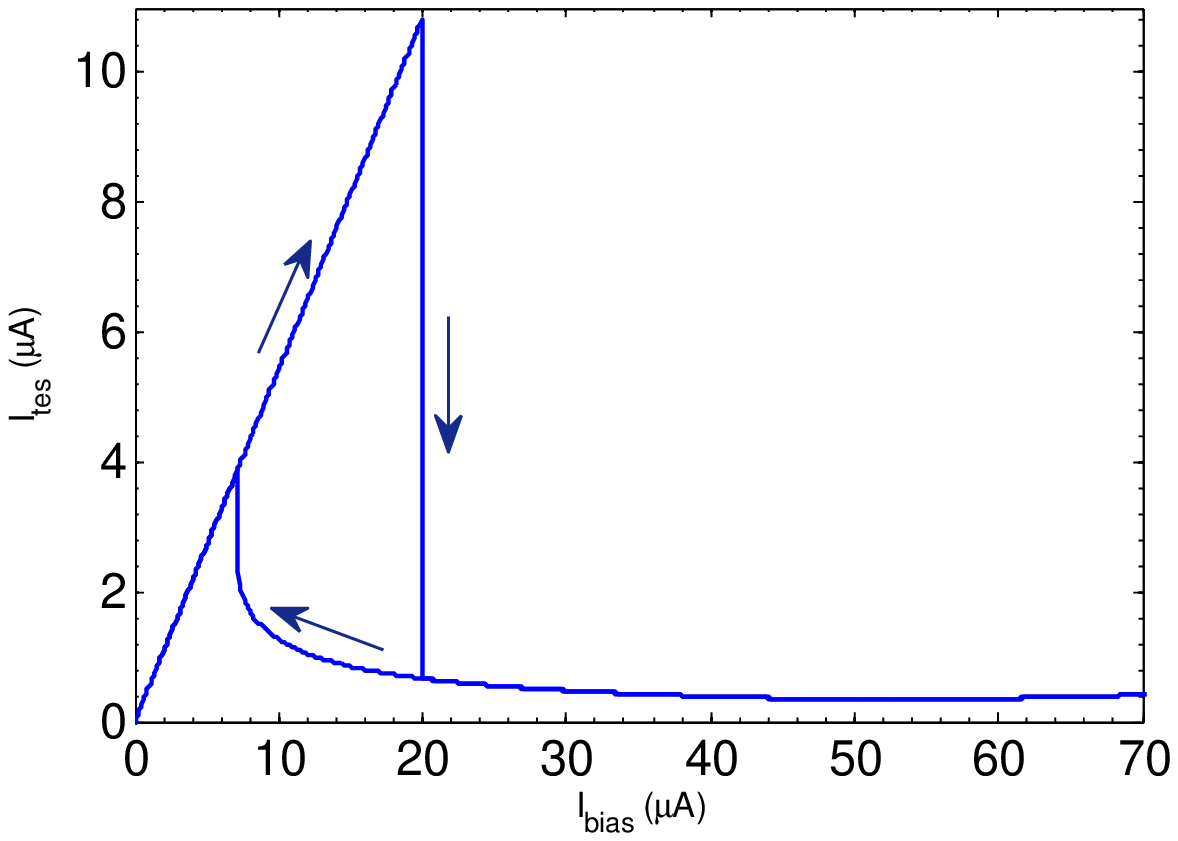}}
  \caption{Measured and simulated TES current against total bias
current for voltage biased TES device.}
  \label{fig:Ttes_Ib}
\end{figure}

Though the temperature-bias current curve cannot be measured
directly to observe the hysteresis in the device temperature,
indirect experimental evidence is available. Some authors have
measured the current in the TES branch against the total bias
current in Fig. \ref{fig:TES_Vbiased}. The data can consist of a
superconducting branch and a resistive branch as shown in Fig.
\ref{fig:Ttes_Ib-experiment} . When the bias current is decreased
from the resistive branch, the TES eventually returns to the
superconducting state, however the bias current at the transition
point is different than that for the superconducting to resistive
transition which leads to a hysteresis structure in Fig.
\ref{fig:Ttes_Ib-experiment}. This is a manifestation of the
hysteresis in the TES resistance, which in turn is due to the
temperature hysteresis in Fig. \ref{fig:T_Ib-1}. The current curve
in Fig. \ref{fig:Ttes_Ib-experiment} can be simulated using our
device model and the result is plotted in Fig.
\ref{fig:Ttes_Ib-simulation} . The result agrees well with
experimental data indicating the effectiveness of the device model.

\subsection{Transient response to signal pulses}
Our device model can be used directly in transient and AC analysis to simulate
temporal and frequency responses of the TES circuits to input signals. The
simulator will automatically linearize the circuit when necessary, saving the
trouble of manually deriving small signal models. \\

\begin{figure}
\centering
\includegraphics[width=2.5in]{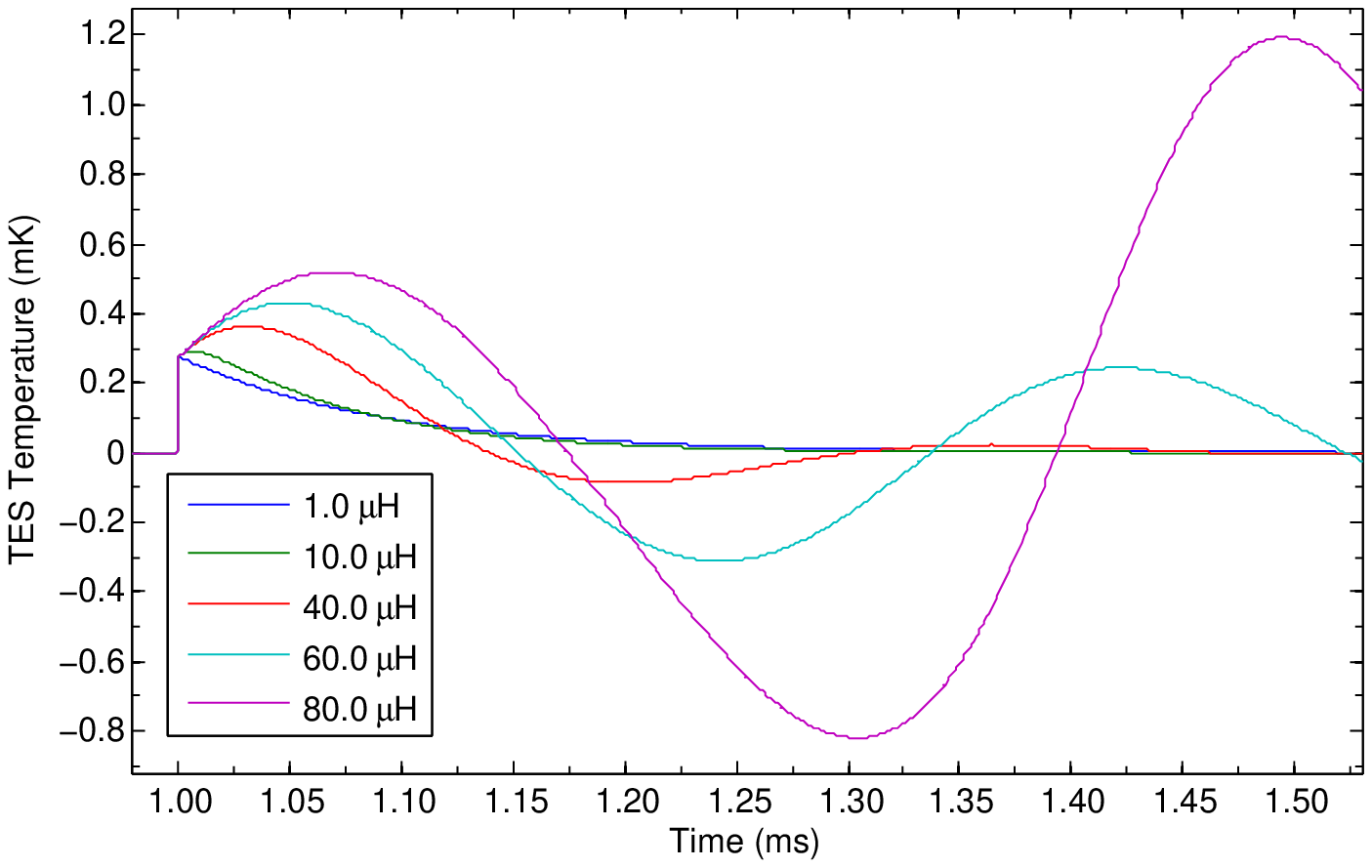}
 \caption{Parametric simulation of the temporal response of the TES circuit in
Fig. \ref{fig:TES_Vbiased}. The input signal is a short pulse. The temperature
of the TES is plotted as a
function of time, for different inductance values. Bias current of
the circuit is $30\mu$A. All other parameters and biases are the
same as in previous simulations (see Fig. \ref{fig:RT_curve}, Fig.
\ref{fig:RT_Vbiased} and Fig. \ref{fig:Ttes_Ib}). }
\label{fig:transient}
\end{figure}

As an example, we simulate the transient response of the TES circuit
in Fig. \ref{fig:TES_Vbiased} to an input signal pulse under
different circuit parameters. The TES temperature as a function of
time is plotted in Fig. \ref{fig:transient}. Using the EDA tool's
parametric analysis functions, we can perform the same simulation
for a range of circuit parameters in just one run and plot the
results in the same figure. This makes it convenient to compare the
results and observe how the response of the circuit changes with
circuit parameters. In Fig. \ref{fig:transient}, we see that the
circuit response becomes unstable when the inductance $L$ increases.
This simple parametric simulation then allows us to determine the
range of acceptable values of the inductance to ensure the stability
of the response (when other circuit parameters are fixed). Such
search for appropriate circuit parameter values is an important task
in circuit design, and it is much more challenging when multiple
parameters need to be considered simultaneously to maximize the
circuit's tolerance to fabrication errors. By developing
sophisticated software that intelligently uses parametric
simulations based on our device model in a multi-dimensional
parameter space, it is possible to automate the critical task of
optimizing circuit parameters \cite{ref:optimization}.

A direct comparison of the simulation results in Fig.
\ref{fig:transient} to experimental data is hindered by the
incompleteness of the device parameters in reference
\cite{ref:Taralli09}. However, the device's temperature sensitivity
suggested by the simulation appears to be smaller than the values
given in the original reference for the same bias current. This
indicates that our TES model based on idealized device physics might
not give completely accurate numerical results for all TES devices
considering the many uncertain and poorly controlled factors in the
fabrication process that can impact the characteristics of the
fabricated device. Of particular importance is the temperature
dependence of the supercurrent in the superconducting to normal
transition region since it determines the sharpness of the
transition and hence the device's temperature sensitivity. It is
conjecturable that the exponent $\lambda$ in the supercurrent -
temperature relation $I_s(T) = I_{s0}(1-T/T_c)^{\lambda}$ can
deviate from the BCS result $\lambda=1.5$ in the transition region
for practical devices, and simulation shows that the device's
temperature sensitivity is very sensitive to the value of $\lambda$.
It is up to further theoretical and experimental studies to
determine whether careful consideration of this issue can explain
the discrepancy between the simulation and experimental data and
lead to more accurate device models.

Though the example simulations we described in this paper are all based on the
simple voltage biased TES circuit in Fig. \ref{fig:TES_Vbiased}, more
sophisticated circuits can be simulated and more complex analysis can be
performed using our device model. If we integrate the device model in a circuit
simulator which supports Josephson devices such as WRspice \cite{ref:wrspice},
we will be able to simulate complete superconducting circuit systems that
contain both TES devices and supporting Josephson circuits (e.g. SQUID
amplifiers and multiplexers). Such powerful tool will make it possible to design
and study large scale TES circuit systems for future scientific applications.

\section{Conclusion}
In summary, we have developed a simple TES device model based on the
superfluid - normal fluid theory. The device model is not limited to
small signal simulations and can be used to study important
characteristics of TES circuits and assist their design. Simulation
results based on our device model are consistent with important
observations and conclusions derived from experimental data, and
they can be used to study phenomena not directly measurable in
experiments. The device model can be improved by
refining the device physics and considering neglected factors such
as magnetic fields and noises. It is hoped that future improved device
models will give better accuracy and reliability, so that they can be
used to develop sophisticated EDA tools that can eventually support the
design and simulation of large scale TES circuits.


\end{document}